\documentclass[%
12pt,T
oneocolumn,
nofootinbib,
amsmath,amssymb,
aps,
floatfix,
unsortedaddress
]{revtex4-2}

\usepackage{graphicx,color}
\usepackage{subfig}
\usepackage{dcolumn}% Align table columns on decimal point
\usepackage{bm}% bold math
\usepackage{graphicx,color}
\usepackage{float}
\usepackage{multirow}
\usepackage{amsmath}
\usepackage{mathtools}
\usepackage{cancel}
\usepackage[pdfpagelabels, pdfencoding=auto, psdextra]{hyperref}
\hypersetup{%
	pdfsubject=Paper,
	pdfkeywords={nuclear physics} {Pionless EFT} {Two-Nucleon} {Lattice} {L\"usher formula},
	unicode = true,
	breaklinks = true,
	colorlinks = true,
	linkcolor = blue,
	citecolor = blue,
	menucolor = blue,
	citecolor = blue,
	urlcolor = blue
}

\graphicspath{{./}}
\begin{document}
	
	\title{Probing $ \phi $N interaction through bound states of $ \phi\textrm{N-}\alpha $ system }
	\author{Faisal Etminan}
	\email{fetminan@birjand.ac.ir}
	\affiliation{
		Department of Physics, Faculty of Sciences, University of Birjand, Birjand 97175-615, Iran
	}%
	\affiliation{ Interdisciplinary Theoretical and Mathematical Sciences Program (iTHEMS), RIKEN, Wako 351-0198, Japan}
	
	\date{\today}% 
	\begin{abstract}
		The possible bound state of the $ \phi\textrm{N-}\alpha $ system is explored within the framework of the  three-body cluster model.
		The calculations are done by employing the state-of-the-art $ \phi $N interactions obtained from the analysis of the pure elastic scattering and the coupled-channel in the $\phi p$ correlation functions.
		The $ \phi\alpha $ potentials are constructed by two methods: the single-folding potential (SFP) method for the given spin-averaged $ \phi $N potentials in coordinate space, and the optical model potential (OMP) approximation within the multiple-scattering framework for the given scattering length of the $ \phi $N interaction. 
		It is found that, when only the single-channel $ \phi $N interactions are employed, the $ \phi\textrm{N-}\alpha $ system could be bound with a binding energy in the interval [3–26] MeV.
		However, the coupled-channel $\phi p$ interaction, which is most consistent with experimental measurements, does not yield any bound state, even when the spin-averaged $\phi$N potential is employed, and this potential is found to be more attractive than the corresponding coupled-channel counterpart.
		It is essential to consider the contributions from the dynamics of the vector-baryon coupled channels $\phi p$ interaction.
		This effect is capable of playing a decisive role in the existence of mesic nuclei.
	\end{abstract}	
	%\begin{keyword}
	%Lattice HAL QCD $ \phi $N interaction, $ \phi\textrm{N-}\alpha $-mesic nuclei, three-body cluster model, hyperspherical harmonics expansions  
	%\end{keyword}	
	\maketitle
	%%%%%%%%%%%%%%%%%%
	\section{Introduction} \label{sec:intro}
	Among the various studies of meson properties in nuclear matter, particular attention has been given to the vector $\phi$-meson, since the interactions of the $s\bar{s}$ pair with nucleons and nuclei are not yet fully understood at low energies. Recently, $\phi$-mesic nuclei—exotic many-body systems characterized by strong interactions—have garnered renewed interest. The coupling of the $\phi$ meson to nucleons suggests that the stable formation of $\phi$-mesic nuclei is probable, and both theoretical models and experimental efforts continue to investigate this possibility~\cite{HIRENZAKI2010406,10.1143/PTP.124.147,PhysRevC.96.035201,PhysRevC.95.055202,etminan2024prc,Filikhin2024prd,filikhin2024phihe,ETMINAN2025PLB139564}. Theoretically, the properties of the $\phi$ meson in nuclear matter have been analyzed based on hadronic models~\cite{PhysRevC.95.015201}, QCD sum rules~\cite{PhysRevD.105.114053}, chiral and effective Lagrangian approaches~\cite{FeijooPRD2025, AbreuPLB2025}.

   A naïve first guess regarding the nature of the $\phi $N interaction would be biased by the OZI (Okubo–Zweig–Iizuka) rule, which would imply a weak interaction between the two hadrons due to their distinct quark contents. A more careful line of reasoning would, however, raise questions about possible coupled-channel effects. 
   Indeed, the experimental study of the $\phi p$ correlation function (CF), measured by the ALICE Collaboration~\cite{PhysRevLett.127.172301}, indicates that the relevant interactions are far from those expected under OZI suppression. A Lednický–Lyuboshits fit to the data yielded a spin-averaged scattering length of
   $ a_{0} = (0.85 \pm 0.48)$ $+$ $\textrm{i}(0.16 \pm 0.19)$ fm.

   Because the imaginary part of \( a_{0} \) is compatible with zero within uncertainties, it is interpreted that elastic $\phi p$ scattering dominates the interaction. With this assumption, a subsequent study~\cite{CHIZZALI2024138358} of the $\phi p$ CF \cite{PhysRevLett.127.172301} employed a single-channel phenomenological potential, fitted to N$\phi$ lattice data by the HAL QCD collaboration~\cite{yan2022prd}, to explore the controversial existence of a $\phi p$ bound state.
    As a result, the binding energy of the $\phi p$ bound state was found within $ \left[12.8 -56.1\right]$ for the spin 1/2 channel. Despite the previous evidence, the nature of this state remains ambiguous. A clear competing interpretation is provided by several studies based on extensions of chiral Lagrangians to accommodate vector mesons within a coupled-channel unitary framework \cite{oset2010dynamically, PhysRevD.84.056017, garzon2012effects, FeijooPRD2025, AbreuPLB2025}.

   Nevertheless, in Ref.~\cite{FeijooPRD2025}, Feijoo et al. present another fit to the same data, in which the CF was calculated using vector-meson–baryon amplitudes obtained within a coupled-channel approach based on hidden local gauge symmetry. In this model, interactions proceed through t-channel vector-meson exchange, which yields spin-independent CFs and differs from the findings of Ref.~\cite{CHIZZALI2024138358}. In Ref.~\cite{FeijooPRD2025}, on the one hand, scattering amplitudes were calculated following~\cite{oset2010dynamically}, constrained by the $\phi p$ femtoscopic data to extract information about the state near the $\phi p$ threshold and the scattering parameters; on the other hand, it is shown that inelastic transitions cannot be underestimated or avoided in the analysis of certain CFs by examining the relative weight of each scattering transition from any member of the coupled channel basis to the measured vector-baryon.

	One of the simplest candidates for $\phi$-mesic nuclei is the $\phi$-NN system~\cite{Belyaev2008,Belyaev2009,Sofianos2010,PhysRevC.108.034614,etminan2024prc,Filikhin2024prd}. Recently, Filikhin et al.~\cite{filikhin2024phihe} studied the $\phi$-mesic nuclei $_{\phi}^{9}\textrm{Be}$ and $_{\phi\phi}^{6}\textrm{He}$ within a three-body cluster model framework, treating them as $\phi\alpha\alpha$ and $\phi\phi\alpha$ systems using the Faddeev formalism in configuration space. The $\phi\alpha$ potential was determined through a folding procedure of the HAL QCD $\phi$N interaction in the $^{{4}}S_{3/2}$ channel~\cite{yan2022prd}, utilizing the matter distribution of $^{{4}}\textrm{He}$~\cite{Krauth2021}.
	
	As mentioned above, a significant difference is observed between the scattering parameters derived from pure elastic single-channel~\cite{CHIZZALI2024138358} and those derived from the coupled-channel~\cite{FeijooPRD2025} $\phi$N interactions; consequently, these differences are sought to be reflected in the structure of the possible bound mesic system.
	Accordingly, the properties of the possible $\phi\textrm{-}N\alpha$ mesic nuclei are examined using $\phi$N potentials in both single- and coupled-channel formulations, with $\alpha$ acting as a spectator, by employing advanced few-body computational techniques.

	In this study, the binding energies and matter radii of $\phi\textrm{-}N\alpha$-mesic system are calculated within a cluster model utilizing the hyperspherical harmonics (HH) expansion method~\cite{Zhukov93,Casal2020,ETMINAN2023122639}. 
	$ \phi\alpha $ potentials are constructed through two methods: the SFP procedure for the spin-averaged $ \phi $N interaction and the OMP approximation within the multiple-scattering framework. Since $ \phi $N potentials in coupled channels are not provided in coordinate space, the OMP approximation is used to build the $ \phi\alpha $ potentials. The optical potential is derived from the projectile–nucleus scattering amplitude, which can be obtained by solving the multiple-scattering equations. The multiple-scattering formalism has been applied to the description of exotic atoms  ($\pi^{-},K^{-},\bar{p}$-atoms) and to resonance-dominated $\pi$-nucleus scattering~\cite{eisenberg_kolton_1988, lenz1987strong}. For N$\alpha$  interactions, two potential forms are considered: the Sack–Bildenharn–Breit (SBB) $V_{n\alpha}$ potential~\cite{Sack-PhysRev.93.321}, characterized by a Gaussian form, and a Woods–Saxon (WS) form used in coordinate-space Faddeev calculations~\cite{BANG1983126,Zhukov93}.
	
	The organization of this paper is as follows. In Section~\ref{sec:Three-body hyperspherical}, a brief outline of the three-body hyperspherical harmonic (HH) approach is provided. The characteristics of the $\phi$N interaction and the choices of $\phi$,N-core interactions are discussed in Section~\ref{sec:Two-body-potentials}. Additionally, the OMP approximation for the $\phi \alpha$ potential, derived from the multiple-scattering formalism, is described.
	 In Section~\ref{result}, numerical results are presented. The final section, Section~\ref{sec:Summary-and-conclusions}, is dedicated to a summary and concluding remarks.
	%%%%%%%%% 
	\section{ Expansion on hyperspherical harmonics }
	\label{sec:Three-body hyperspherical}
	The expansion on HH method is developed version of the Faddeev equations in coordinate space.
	Since this method has been explained in details in other works~\cite{Zhukov93,raynal1970,face}, I shall here give a rather brief description
	of this formalism.
	
\begin{figure*}[hbt!]
		\centering
		\includegraphics[scale=0.15]{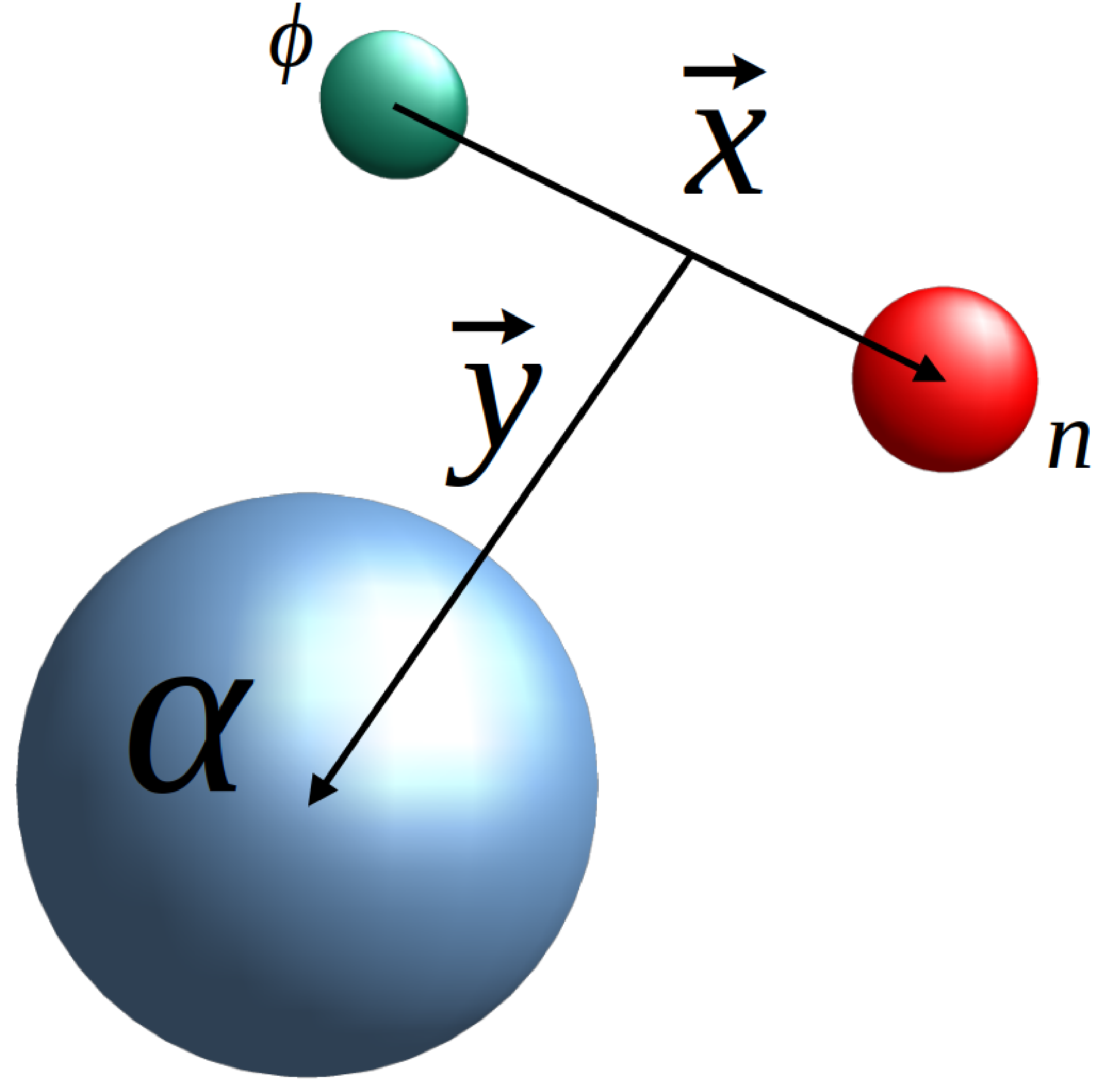}
		\caption{The Jacobi T-coordinate system used to describe the $ _{\phi}^{6}\textrm{He} $ mesic nuclei as $\phi\textrm{N-}\alpha$. }
		\label{fig:T_jacobi}
	\end{figure*}

	In a three-body $\phi $+N+core model, the total three-body wave function $\Psi$ is sum of three components, i.e.,
	the (usually) inactive core intrinsic wave function and the active part that 
	it depends on the relative coordinates and spins (often suppressed) of the two valence particles, 
	$\Psi^{(i)}$, i.e, $\Psi=\sum_{i=1}^{3}\Psi^{\left(i\right)}$. 
	The components $\Psi^{\left(i\right)}$ are function of the three different sets of
	Jacobi coordinates, One of the three sets is shown in the Fig.~\ref{fig:T_jacobi}, and they satisfy the three Faddeev coupled equations,
	\begin{equation}
		\left(T-E\right)\Psi^{\left(i\right)}+V_{jk}\left(\Psi^{\left(i\right)}+\Psi^{\left(j\right)}+\Psi^{\left(k\right)}\right)=0,
		\label{eq:faddeev_coupled-eq}
	\end{equation}
	where $T$ is the kinetic energy, $E$ is the total energy, $V_{jk}\left(r_{jk}\right)$ is
	the two body interactions between the corresponding pair and the indexes
	$i,j,k$ is a cyclic permutation of $\left(1,2,3\right)$.
	
	It is convenient to introduce translation invariant sets of Jacobi coordinates, 
	$ \left\{ \vec{x},\text{\ensuremath{\vec{y}}}\right\} $ 
	illustrated in Fig.~\ref{fig:T_jacobi}, are employed to describe three-body systems [six-dimensional problems].
	The variable $ \vec{x} $ represents the relative coordinates between two of the particles and $ \vec{y} $ is between their center of mass and the third particle,  $\vec{x_{i}}=\sqrt{A_{jk}}\vec{r}_{jk}$ and $\vec{y_{i}}=\sqrt{A_{\left(jk\right)i}}\vec{r}_{\left(jk\right)i}$ where
	$\vec{r}_{jk}=\vec{r}_{j}-\vec{r}_{k}, \vec{r}_{\left(jk\right)i}=\vec{r}_{i}-\left(A_{j}\vec{r}_{j}+A_{k}\vec{r}_{k}\right)/\left(A_{j}+A_{k}\right)$
	with the scaling mass factor are defined by 
	\begin{equation}
		\begin{array}{cc}
			A_{jk}=\frac{A_{j}A_{k}}{A_{j}+A_{k}}, & A_{\left(jk\right)i}=\frac{\left(A_{j}+A_{k}\right)A_{i}}{A_{i}+A_{j}+A_{k}},\end{array}
	\end{equation}
	where $ A_{i}=m_{i}/m $ with $ m=1 $ a.m.u. and $ m_{i} $ the mass of particle $i$ in a.m.u.
	The Jacobi coordinates are transformed into the hyperspherical coordinates $\{ \rho, \alpha,\Omega_{x},\Omega_{y} \} $, 
	with hyperradius $\rho^{2}=x^{2}+y^{2}$ and the hyperangle  $\alpha=\arctan (x/y )$.
	The angles $\Omega_{x}$ and $\Omega_{y}$, which define the directions of the vectors $\vec{x}$ and $\vec{y}$ respectively, are described. For simplicity, all angular dependencies are expressed through $\varphi = (\alpha, \Omega_{x}, \Omega_{y})$.
	%%%%%%%%%%%
	%%%%%%%%%%%
	The Hamiltonian of a three-body system in hyperspherical coordinates is given by
	\begin{equation}
		\hat{H}=\hat{T}\left(\rho,\varphi\right)+\hat{V}\left(\rho,\varphi\right), \label{eq:Hamiltonian}
	\end{equation}
	where the potential operator $\hat{V}(\rho, \varphi)$, representing the sum of pairwise interactions, is included, and the free Hamiltonian operator $\hat{T}(\rho, \varphi)$ is defined as in Ref.~\cite{lay2012}
	\begin{equation}
		\hat{T}(\rho, \varphi) = -\frac{\hbar^{2}}{2m} \left( \frac{\partial^{2}}{\partial\rho^{2}} + \frac{5}{\rho} \frac{\partial}{\partial\rho} - \frac{1}{\rho^{2}} \hat{K}^{2}(\varphi) \right),
	\end{equation}
	where the hyperangular momentum operator, $\hat{K}$, which is regarded as a generalized angular momentum operator, is introduced. The mass $m$, chosen as the normalization mass, is noted to be $m = m_{N}$.

	Solutions to the Schr\"odinger equation with the three-body Hamiltonian given in Eq.~\eqref{eq:Hamiltonian} can be expanded for each $\rho$, with the total angular momentum $j$ considered, as
	\begin{equation}
		\psi_{i\beta}^{j\mu}(\rho,\varphi)=R_{i\beta}(\rho)\mathcal{Y}_{\beta}^{j\mu}(\varphi),
	\end{equation}
	where the functions $\mathcal{Y}_{\beta}^{j\mu}(\varphi)$ form a complete set of hyperangular functions. These functions are expanded in hyperspherical harmonics as presented in Refs.~\cite{Zhukov93,Casal2020,raynal1970}, and the hyperradial functions $R_{i\beta}(\rho)$ are introduced, with the subscript $i$ indicating the hyperradial excitation. To facilitate the solution of the coupled equations~\eqref{eq:Faddeev_coupled}, the hyperradial functions, denoted as $\mathcal{R}_{\beta}^{j}(\rho)$, are expanded in an orthonormal discrete basis up to $i_{\max}$~\cite{face}.
	
	The set of quantum numbers $\beta \equiv \{K, l_x, l_y, l, S_x, j_{ab}\}$ is used, where $l_x$ and $l_y$ are the orbital angular momenta associated with the Jacobi coordinates $\vec{x}$ and $\vec{y}$ respectively. The total orbital angular momentum $l$ is given by the sum $l_x + l_y$, and the total spin of the particle pair linked via $\vec{x}$, $S_x$, is included. The quantum number $j_{ab} = l + S_x$ is used, and the total angular momentum $j$ is expressed as $j = j_{ab} + I$, with $I$ indicating the spin of the third particle.
	
	The wave functions of the system are defined as 
	\begin{eqnarray}
		\Psi^{j\mu}(\rho,\varphi) & = & \sum_{\beta}\sum_{i=0}^{i_{max}}C_{i\beta}^{j}\:\psi_{i\beta}^{j\mu}(\rho,\varphi)\\
		& = & \sum_{\beta}\left(\sum_{i=0}^{i_{max}}C_{i\beta}^{j}R_{i\beta}(\rho)\right)\mathcal{Y}_{\beta}^{j\mu}(\varphi)=\sum_{\beta}\mathcal{R}_{\beta}^{j}(\rho)\mathcal{Y}_{\beta}^{j\mu}(\varphi),\nonumber 
	\end{eqnarray}
	where the coefficients $C_{i\beta}^{j}$ are obtained through diagonalization of the three-body Hamiltonian within the basis functions up to $i_{\max}$. The hyperradial wave functions $\mathcal{R}_{\beta}(\rho)$ are solutions to the coupled differential equations, expressed as
	\begin{equation}
		\left(-\frac{\hbar^{2}}{2m}\left(\frac{d^{2}}{d\rho^{2}}-\frac{(K+3/2)(K+5/2)}{\rho^{2}}\right)-E\right)\mathcal{R}_{\beta}^{j}(\rho)+\sum_{\beta'}V_{\beta'\beta}^{j\mu}(\rho)\mathcal{R}_{\beta'}^{j}(\rho)=0,
		\label{eq:Faddeev_coupled}
	\end{equation} 
	where the term $V_{\beta'\beta}^{j\mu}(\rho)$, associated with two-body potentials between pairs of particles ($V_{ij}$), is defined by
	\begin{equation}
		V_{\beta'\beta}^{j\mu}(\rho)=\left\langle \mathcal{Y}_{\beta}^{j\mu}(\varphi)\left|V_{12}+V_{13}+V_{23}\right|\mathcal{Y}_{\beta^{\prime}}^{j\mu}(\varphi)\right\rangle. 
	\end{equation}
	
	Calling $r$ the modulus of the coordinate relating two particles, the general binary interaction potential, involving central, spin-orbit, spin-spin, and tensor components, is characterized by
	\begin{equation}
		\hat{V}_{ij}=V_{c}\left(r\right)+V_{so}\left(r\right)\vec{l}_{ij}.\vec{s}_{ij}+V_{ss}\left(r\right)\vec{s}_{i}.\vec{s}_{j}+V_{T}\left(r\right)\hat{S}_{ij},
	\end{equation}
	where $\vec{s}_i$ denotes the spin of particle $i$, and $\vec{s}_{ij} \equiv \vec{s}_i + \vec{s}_j$. The relative orbital angular momentum between particles $i$ and $j$ is represented by $\vec{l}_{ij}$, while $\hat{S}_{ij}$ stands for the tensor operator. The radial form factors for each term may be chosen from functions such as Gaussian, exponential, Woods-Saxon, or Yukawa functions.
	%%%%%%%%
	\section{Choice of $ \phi $N  and $ \textrm{N} $-core interactions }
	\label{sec:Two-body-potentials}
	The unique aspect of the three-body problem lies in its sensitivity to the off-shell behavior of the two-body interaction. To examine this, various choices of $\phi$N and N$\alpha$ interactions were employed in calculations of $\phi\textrm{N-}\alpha$ system. 
	
	%%%%%%%%%%
	\subsection{ A single channel $ \phi $N potentials from lattice QCD}
	As reported in Ref.~\cite{yan2022prd}, the $\phi \textrm{N}\left(^{4}S_{3/2}\right)$ potential is characterized by a combination of an attractive core at short distances and a two-pion-exchange (TPE) tail at long distances, with an overall strength proportional to $m_{\pi}^{4}$. Utilizing the HAL QCD method, it was found that the $\phi$N correlation function is predominantly influenced by elastic scattering states in the $^{4}S_{3/2}$ channel, with no significant contributions from the two-body channels $\Lambda K\left(^{2}D_{3/2}\right)$ and $\Sigma K\left(^{2}D_{3/2}\right)$, nor from three-body open channels such as $\phi \textrm{N} \rightarrow \Sigma^{*}K$, as well as those involving $\Lambda(1405)K \rightarrow \Lambda\pi K, \Sigma\pi K$. Consequently, the effects arising from these coupled channels are not visible in lattice calculations. The $\phi$N interaction in the $^{2}S_{1/2}$ state differs since it can couple via S-wave to the $\Lambda K\left(^{2}S_{1/2}\right)$ and $\Sigma K\left(^{2}S_{1/2}\right)$ states~\cite{yan2022prd}. Such decay processes are described through kaon exchange, which introduces an imaginary component to the potential in the $^{2}S_{1/2}$ channel.
	
	Conversely, the long-range TPE potential observed in the $^{4}S_{3/2}$ channel is expected to also characterize the $^{2}S_{1/2}$ channel. This is because the exchange of two pions in a scalar-isoscalar state does not depend on the total spin of the $\phi$N system. 
	
	Based on the results presented in Refs.~\cite{PhysRevLett.127.172301, yan2022prd}, a fit was performed to experimental correlation data~\cite{CHIZZALI2024138358}, constraining the spin $3/2$ interaction using the scattering length obtained from lattice QCD simulations~\cite{yan2022prd}. These considerations lead to the proposed potential form for the $\phi$N system in the $^{2}S_{1/2}$ channel~\cite{CHIZZALI2024138358}:
	\begin{equation}
		V_{\phi \textrm{N}}\left(r\right)=\beta\left(\sum_{i=1}^{2}a_{i}e^{-\left(r/b_{i}\right)^{2}}\right)+a_{3}m_{\pi}^{4}f\left(r;b_{3}\right)\left(\frac{e^{-m_{\pi}r}}{r}\right)^{2}+i\gamma f\left(r;b_{3}\right)\frac{e^{-2m_{K}r}}{m_{K}r^{2}}, \label{eq:phiN-pot}
	\end{equation}
	where $\beta$ and $\gamma$ are parameters adjusted to fit experimental data; $a_{i}$ and $b_{i}$ are common to both the $^{4}S_{3/2}$ and $^{2}S_{1/2}$ channels, obtained through fits to lattice QCD data in the spin $3/2$ channel at $t/a=14$~\cite{yan2022prd} and $t/a=12$~\cite{CHIZZALI2024138358}, respectively. These parameter values are listed in Table~\ref{tab:Fit_para}. Here, $t/a$ represents Euclidean (imaginary) time with a lattice spacing of $a=0.0846$ fm.
	
	In fact, in Ref.~\cite{CHIZZALI2024138358}, the $t/a=12$ value was chosen over the default $t/a=14$ used in~\cite{yan2022prd} to adopt a conservative approach, as it corresponds to the least attractive potential in the spin $3/2$ channel, and consequently, to the weakest potential in the spin $1/2$ channel. The function $f\left(r;b_{3}\right)=\left(1-e^{-\left(r/b_{3}\right)^{2}}\right)^{2}$ is an Argonne-type form factor, discussed in~\cite{yan2022prd}.
	
	For $\beta=1$ and $\gamma=0$, the potential in Eq.~\eqref{eq:phiN-pot} reduces to the lattice QCD potential for the spin $3/2$ channel~\cite{yan2022prd}. Conversely, the fitted parameters $\beta=6.9_{-0.5}^{+0.9}\left(\textrm{stat.}\right)_{-0.1}^{+0.2}\left(\textrm{syst.}\right)$ and  $\gamma=0.0_{-3.6}^{+0.0}\left(\textrm{stat.}\right)_{-1.8}^{+0.0}\left(\textrm{syst.}\right)$ are extracted through fitting to the experimental p$\phi$ correlation data measured by the ALICE collaboration in pp collisions at $\sqrt{s}=13$ TeV~\cite{PhysRevLett.127.172301}.
	
	In addition, in Ref.~\cite{CHIZZALI2024138358}, physical masses were employed for both $m_{\pi}=138$ MeV/$c^{2}$ and $m_{K}=496$ MeV/$c^{2}$ to accurately model the experimental correlation function; similarly, the physical masses are used here.
	The HAL QCD $\phi$N two-body potential in the $^{4}S_{3/2}$ channel (as given in Eq.~\eqref{eq:phiN-pot} with parameters listed in Table~\ref{tab:Fit_para}) does not produce a bound state. However, the $\phi$N system in the $^{2}S_{1/2}$ channel is strongly bound, with a central binding energy of $B_{\phi \textrm{N}}=28.93$ MeV at $t/a=12$~\cite{CHIZZALI2024138358}, and $16.64$ MeV at $t/a=14$~\cite{yan2022prd}.
	%%%%%%%%%%%%%%%%
	\begin{table}[htp]
		\caption{The parameters of  $ \phi $N  potential given in Eq.~\eqref{eq:phiN-pot} at lattice Euclidean time $ 12 $~\cite{CHIZZALI2024138358} and $ 14 $~\cite{yan2022prd}. The numbers in  parentheses indicate statistical errors.}
		\begin{tabular}{ccccccc}
			\hline
			$ t/a $	&$a_1~(\mathrm{MeV})$ &	$b_1~(\mathrm{fm})$ &	$a_2~(\mathrm{MeV})$ & $b_2~(\mathrm{fm})$ & $a_3 m_{\pi}^{4} ~(\mathrm{MeV \cdot fm^{2}})$ & $b_{3}(\mathrm{fm})$ \\
			\hline  
			$ 12 $&$ -392(10)$  &$ 0.128(3) $ &$ -145(9) $  & $ 0.284(7) $ & $ -83(1) $  & $ 0.582(6) $  \\
			\hline
			$ 14 $&$ -371(27$)  &$ 0.13(1) $  &$ -119(39) $  & $ 0.30(5) $ & $ -97(14) $  & $ 0.63(4) $  \\
			\hline
		\end{tabular}
		\label{tab:Fit_para}
	\end{table}
	%%%%%%%%%%%%%%	
	
\begin{figure*}[hbt!]
		\centering
		\includegraphics[scale=1.0]{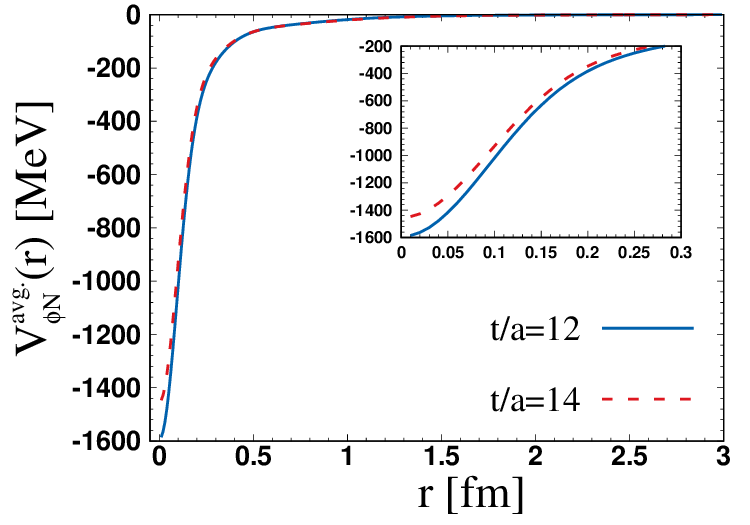}
		\caption{The spin-averaged $ \phi $N potential in Eq.~\eqref{eq:spin-ave}, as a function of separation $ r $ for  lattice Euclidean times $t/a= 12 $ (blue solid line), and $ 14 $ (red dashed lines) by the parameters given in Table~\ref{tab:Fit_para}.}
		\label{fig:Nphi_pot}
	\end{figure*}

	In Fig.~\ref{fig:Nphi_pot}, the spin-averaged $\phi$N potentials, as defined by Eq.~\eqref{eq:spin-ave}, are displayed at time slices $t/a=12$ and $14$ using the parameters listed in Table~\ref{tab:Fit_para} to facilitate better comparison.
	%%%%%%%%%%%%%%	
	%%%%%%%%%%%
	\subsection{A coupled channel $ \phi $N interaction from chiral Lagrangians }	
 Feijoo et al. \cite{FeijooPRD2025} examined the validity of their approach with two models, a purely theoretical model and a bootstrap model, in which, in the latter, the parameters and their uncertainties are obtained (as well as every other quantity, e.g., scattering lengths) via bootstrap techniques \cite{efron1986bootstrap} using the data points available in the $\phi p$ femtoscopic region \cite{PhysRevLett.127.172301}. 
 The $\phi$p effective range and scattering length extracted from the models are 
  $ \left(a_{0}[\textrm{fm}],r_{0}[\textrm{fm}]\right) =  \left(0.272 + \textrm{i} 0.189, -7.20 - \textrm{i} 0.09\right)$ and $\left(-0.034 + \textrm{i} 0.57,-8.06 + \textrm{i} 0.05\right)$.

 The new results differ significantly from previous results \cite{PhysRevLett.127.172301, CHIZZALI2024138358}. Moreover, it is shown that the dynamics of the vector-baryon coupled-channel interaction is essential to deduce the $\phi$N interaction, which is neglected in the lattice QCD simulation of the $\phi$N interaction \cite{yan2022prd}. Therefore, an optical potential for the $\phi \alpha$ interaction is presented in the following within the multiple-scattering formalism, employing the scattering lengths obtained from pure theoretical  models, and these results are compared with those based on lattice simulations of the $\phi$N interaction.

	\subsection{Construction of $\phi\alpha$ potential}
	\subsubsection{SFP method}
	The spin-averaged WS type form of the potential,
	\begin{equation}
		U_{\phi\alpha}^{fit}\left(r\right)=-U_{0}\left[1+\exp\left(\frac{r-R_{\phi\alpha}}{t_{\phi\alpha}}\right)\right]^{-1},  \label{eq:ws-fit}
	\end{equation}
	for the $\phi \alpha$ potential, $U_{\phi\alpha}$, is obtained within the framework of the SFP approach. In this approach, the effective $\phi \alpha$ nuclear potential is estimated through the single folding of the nucleon density in the $\alpha$-particle with the spin-averaged $\phi$N interaction~\cite{Satchler1979,10.1143/ptp.117.251,Etminan:2019gds, filikhin2024phihe}. The spin-averaged $\phi$N potential is defined as 
	\begin{equation}
		V_{\phi N}^{avg.}\left(r\right)=\frac{2}{3}V_{\phi N}\left(^{4}S_{3/2}\right)+\frac{1}{3}V_{\phi N}\left(^{2}S_{1/2}\right), \label{eq:spin-ave}
	\end{equation}
	where the potential in the $^{2}S_{1/2}$ channel is notably more attractive in the short-range part due to the enhanced $\beta$, and the long-range tail is similar to that of the $^{4}S_{3/2}$ channel, arising from two-pion exchange~\cite{CHIZZALI2024138358}. It should be noted that, in Ref.~\cite{filikhin2024phihe}, the $U_{\phi\alpha}$ potentials were constructed solely by folding the $\phi$N potential in the $^{4}S_{3/2}$ channel. To achieve higher computational accuracy, such an approach is not adopted here.
	
	Recently, the root-mean-square (rms) matter radius of $^4$He was measured to be $1.70 \pm 0.14$ fm through analyses of near-threshold $\phi$-meson photoproduction data from the LEPS Collaboration~\cite{PhysRevC.97.035208, PhysRevC.109.L012201}. These studies indicate that the rms charge radius of $^4$He is smaller than the rms matter radius; however, the two quantities are statistically compatible within their error margins. Despite this apparent discrepancy, the densities used in the present folding potential calculations are modeled to reproduce an rms radius of 1.70 fm.

	\subsubsection{Optical model potential approximation from multiple-scattering formalism}
	A description of the main aspects of the multiple-scattering formalism is provided for the present purposes, as given in the ~\cite{eisenberg_kolton_1988, Lenz1993}. The impulse approximation underpins this formalism, in which the scattering of a projectile by a bound nucleon is approximately identified with scattering on a single free heavy nucleon. Projectile–nucleus scattering is subsequently described as a sequence of such quasi-free scattering events. The bound projectile–nucleon $t$-matrix is defined by the integral equation
	\begin{equation}
		\tau_{i}=v_{i}+v_{i}G_{0}\tau_{i},
	\end{equation}
	where \(v_{i}\) denotes the interaction between the projectile and nucleon \(i\), and
	\(G_{0}=\left(E-H_{0}+\mathrm{i}\eta\right)^{-1}\) represents the Green’s function of the projectile–nucleus system in the absence of projectile–nucleon interactions, with the outgoing-wave boundary condition implemented via \(\eta>0\). The quantity \(\tau_{i}\) is related to the free projectile–nucleon $t$-matrix by
	\begin{equation}
		\tau_{i}=\left(1-v_{i}G_{0}\right)^{-1}v_{i}.
	\end{equation}
	
	Because of the presence of the many-body Green’s function \(G_{0}\), \(\tau_{i}\) is a complicated many-body operator that incorporates the influence of binding forces on the projectile–nucleon scattering. The Lippmann–Schwinger equation in the multiple-scattering formulation is expressible as
	\begin{equation}
	T_{i}=\tau_{i}+\tau_{i}G_{0}\sum_{j\ne i}T_{j},\label{eq:multiple-scatt}
\end{equation}
	where \(T_{i}\) describes a scattering event in which the last scattering occurred with nucleon \(i\). In effect, in the above equation the interaction with projectile \(i\) has been incorporated into the bound transition operator \(\tau_{i}\). The total transition operator \(T=\sum_{i}^{A}T_{i}\) acts on both the projectile and nuclear degrees of freedom, and specific transition amplitudes are obtained by projection onto definite projectile plane-wave states and nuclear initial and final states. In particular, elastic scattering is represented by the matrix element
	\begin{equation}
		T_{0}\left(\boldsymbol{k}^{\prime},\boldsymbol{k}\right)\equiv\left\langle 0,\boldsymbol{k}^{\prime}\left|T\right|\boldsymbol{k},0\right\rangle ,\label{eq:elastic-scatt}
	\end{equation}
	where the incident and outgoing momenta are \(\boldsymbol{k}\) and \(\boldsymbol{k}^{\prime}\), respectively, and the nuclear ground state is \(|0\rangle\). The fixed-scatterer approximation is the basis for most applications of the multiple-scattering formalism; within this approximation the complicated many-body transition operator \(\tau_{i}\) is replaced by a nontrivial operator that primarily enacts momentum conservation and acts on the projectile states. Its momentum-space matrix elements can be written as
	\begin{equation}
		\left\langle \boldsymbol{k}^{\prime},\boldsymbol{p}_{i}^{\prime}\left|\tau_{i}\right|\boldsymbol{k},\boldsymbol{p}_{i}\right\rangle \approx\tau_{i}\left(E,\boldsymbol{k}^{\prime}-\boldsymbol{k}\right)\delta\left[\boldsymbol{k}^{\prime}+\boldsymbol{p}_{i}^{\prime}-\left(\boldsymbol{k}+\boldsymbol{p}_{i}\right)\right],\label{eq:approximation}
	\end{equation}
	where \(\boldsymbol{p}_{i}\) and \(\boldsymbol{p}_{i}^{\prime}\) denote the initial and final momenta of nucleon \(i\). The function \(\tau_{i}\left(E,\boldsymbol{k}^{\prime}-\boldsymbol{k}\right)\) is generally identified with the free projectile–nucleon amplitude, and it is assumed for simplicity that this amplitude depends only on the momentum transfer. This enables the definition of operators that act solely on the projectile degrees of freedom while depending parametrically on the nucleon coordinates. Configuration-space matrix elements are first determined by Fourier transforming
	\begin{equation}
		\left\langle \boldsymbol{r}^{\prime},\boldsymbol{r}_{i}^{\prime}\left|\tau_{i}\right|\boldsymbol{r},\boldsymbol{r}_{i}\right\rangle =\tilde{\tau}_{i}\left(\boldsymbol{r}-\boldsymbol{r}_{i}\right)\delta\left(\boldsymbol{r}-\boldsymbol{r}^{\prime}\right)\delta\left(\boldsymbol{r}_{i}^{\prime}-\boldsymbol{r}_{i}\right),\label{eq:approximation-1}
	\end{equation}
	with
	\begin{equation}
		\tilde{\tau}_{i}\left(\boldsymbol{r}-\boldsymbol{r}_{i}\right)=\int d^{3}q\exp\left[i\boldsymbol{q}\left(\boldsymbol{r}-\boldsymbol{r}_{i}\right)\right]\tau_{i}\left(E,\boldsymbol{q}\right),
	\end{equation}
	and the local operator \(\hat{\tau}_{i}\) is then determined by
	\begin{equation}
		\left\langle \boldsymbol{r}^{\prime}\left|\hat{\tau}_{i}\left(\boldsymbol{r}_{i}\right)\right|\boldsymbol{r}\right\rangle =\delta\left(\boldsymbol{r}-\boldsymbol{r}^{\prime}\right)\tilde{\tau_{i}}\delta\left(\boldsymbol{r}-\boldsymbol{r}_{i}\right).\label{eq:conf_space}
	\end{equation}
	
	Projection of the multiple-scattering Eq.~\eqref{eq:multiple-scatt}, onto nuclear states with the nucleons fixed at positions \(\boldsymbol{r}_{i}\) yields the following operator equation in the Hilbert space of projectile states:
	\begin{equation}
		\hat{T}_{i}=\hat{\tau}_{i}+\hat{\tau}_{i}\frac{1}{E-K+\textrm{i}\eta}\sum_{j\ne i}^{A}\hat{T}_{j},\label{eq:pn_scatt_amp}
	\end{equation}
	where
\begin{equation}
		\left\langle \boldsymbol{r}_{1}^{\prime},\boldsymbol{r}_{2}^{\prime},...,\boldsymbol{r}_{A}^{\prime}\left|T_{i}\right|\boldsymbol{r}_{1}, ...,\boldsymbol{r}_{A}\right\rangle =\hat{T}_{i}\left(\boldsymbol{r}_{1}, ...,\boldsymbol{r}_{A}\right)\prod_{l=1}^{A}\delta\left(\boldsymbol{r}_{l}-\boldsymbol{r}_{l}^{\prime}\right).
\end{equation}
	
	According to the Eqs.~\eqref{eq:multiple-scatt} and~\eqref{eq:elastic-scatt}, the projectile–nucleus elastic amplitude is given by
	\begin{equation}
		T_{0}\left(\boldsymbol{k},\boldsymbol{k}^{\prime}\right)=\sum_{i=1}^{A}\int \textrm{d}^{3}r_{1}, ..., \textrm{d}^{3}r_{A}\:\rho\left(\boldsymbol{r}_{1}, ...,\boldsymbol{r}_{A}\right)\left\langle \boldsymbol{k}^{\prime}\left|\hat{T}_{i}\left(\boldsymbol{r}_{1}, ...,\boldsymbol{r}_{A}\right)\right|\boldsymbol{k}\right\rangle ,\label{eq:elastic-scatt-1}
	\end{equation}
	and the nuclear ground-state density is written as the square of the wave function,
	\begin{equation}
		\rho\left(\boldsymbol{r}_{1}, ...,\boldsymbol{r}_{A}\right)=\left|\left\langle \left.\boldsymbol{r}_{1}, ...,\boldsymbol{r}_{A}\right|0\right\rangle \right|^{2}.\label{eq:gs-density}
	\end{equation}
	Thus, in the fixed-scatterer approximation, the projectile–nucleus scattering amplitude is obtained by solving the multiple-scattering equation (Eq.~\eqref{eq:pn_scatt_amp}) for an arbitrary configuration of nucleons and by configurational averaging of these amplitudes with the weight given by the ground-state density in Eq.~\eqref{eq:approximation-1}.

   Although the quite severe approximations already necessary for deriving Eq.~\eqref{eq:pn_scatt_amp} are present, the amplitude in Eq.~\eqref{eq:approximation-1} is not generally evaluated exactly. Rather, these equations are employed as starting points for further approximations. In the most commonly used approximation, the so-called optical-potential approximation, the amplitudes for the individual configurations are not averaged; instead, the amplitude is calculated for an averaged configuration. The projection of Eq.~\eqref{eq:pn_scatt_amp} onto the nuclear ground-state wave function and the approximation
	$$ 		\sum_{i=1}^{A}\int \textrm{d}^{3}r_{1}, ..., \textrm{d}^{3}r_{A}\:\rho\left(\boldsymbol{r}_{1}, ...,\boldsymbol{r}_{A}\right)\hat{\tau}_{i}\left(\boldsymbol{r}_{i}\right)\frac{1}{E-K+\textrm{i}\eta}\sum_{j\ne i}^{A}\hat{T}_{j}\left(\boldsymbol{r}_{1}, ...,\boldsymbol{r}_{A}\right)$$
	\begin{equation}
\approx U\left(\boldsymbol{r}\right)\frac{1}{E-K+\textrm{i}\eta}\bar{T},\label{eq:wave-appr}
	\end{equation}
where the transition operator \(\bar{T}\) and the optical-potential operator \(U\) are defined by
	\begin{equation}
		\bar{T}=\sum_{i=1}^{A}\int \textrm{d}^{3}r_{1}, ..., \textrm{d}^{3}r_{A}\:\hat{T}_{i}\left(\boldsymbol{r}_{1}, ...,\boldsymbol{r}_{A}\right)\rho\left(\boldsymbol{r}_{1}, ...,\boldsymbol{r}_{A}\right),
	\end{equation}
	\begin{equation}
		U\left(\boldsymbol{r}\right)=\frac{A-1}{A}\sum_{i=1}^{A}\int \textrm{d}^{3}r_{1}, ..., \textrm{d}^{3}r_{A}\:\rho\left(\boldsymbol{r}_{1}, ...,\boldsymbol{r}_{A}\right)\hat{\tau}_{i}\left(\boldsymbol{r}_{i}\right),\label{eq:optical_pot}
	\end{equation}
  For elastic scattering, \(\bar{T}\) satisfies a standard one-body Lippmann–Schwinger equation
	equation
	\begin{equation}
		\bar{T}=\frac{A}{A-1}U+U\frac{1}{E-K+\textrm{i}\eta}\bar{T},
	\end{equation}
	In configuration space, \(U\) from Eq. \eqref{eq:conf_space} is local and is obtained by folding the ground-state density with the projectile–nucleon amplitude:
	\begin{equation}
		\left\langle \boldsymbol{r}^{\prime}\left|U\right|\boldsymbol{r}\right\rangle =\frac{A-1}{A}\delta\left(\boldsymbol{r}-\boldsymbol{r}^{\prime}\right)\sum_{i=1}^{A}\int \textrm{d}^{3}r_{1}, ..., \textrm{d}^{3}r_{A}\:\rho\left(\boldsymbol{r}_{1}, ...,\boldsymbol{r}_{A}\right)\tilde{\tau}_{i}\left(\boldsymbol{r}-\boldsymbol{r}_{i}\right).\label{eq:optical_pot_in_space}
	\end{equation}
	
	In the approximation made by Eq. \eqref{eq:wave-appr}, only the nuclear ground state appears as an intermediate state between two scatterings; hence the optical potential in Eq. \eqref{eq:optical_pot} is the ground-state average of the potentials corresponding to individual configurations. Assuming a zero-range projectile–nucleon \(t\)-matrix, Eq. \eqref{eq:conf_space} becomes
	\begin{equation}
		\tilde{\tau}_{i}\left(\boldsymbol{r}-\boldsymbol{r}_{i}\right)=-\frac{4\pi\hbar^{2}}{2m}a_{0}\delta\left(\boldsymbol{r}-\boldsymbol{r}_{i}\right),\label{eq:zero-range-pn_t-mat}
	\end{equation}
	where \(a_{0}=a_{0}\left(E\right)\) is the projectile–nucleon isotropic scattering amplitude. According to Eq. \eqref{eq:optical_pot_in_space} and the zero-range parametrization, the optical potential is
	\begin{equation}
		U\left(r\right)=4\pi a_{0}\left(A-1\right)\rho\left(r\right). \label{eq:U_rho}
	\end{equation}
	For \({}^{4}\)He, with the center of mass fixed at the origin, the ground-state density \(\rho\left(\boldsymbol{r}_{1},\boldsymbol{r}_{2},\boldsymbol{r}_{3}\right)\) is a function of three independent position vectors and is related to the ground-state density defined in Eq. \eqref{eq:gs-density} by
	\begin{equation}
		\rho\left(\boldsymbol{r}_{1},\boldsymbol{r}_{2},\boldsymbol{r}_{3},\boldsymbol{r}_{4}\right)=\delta\left(\sum_{i=1}^{4}\boldsymbol{r}_{i}\right)\rho\left(\boldsymbol{r}_{1},\boldsymbol{r}_{2},\boldsymbol{r}_{3}\right).\label{eq:gs-3r}
	\end{equation}
	The calculations are performed for the following ground-state density:
	\begin{equation}
		\rho\left(\boldsymbol{r}_{1},\boldsymbol{r}_{2},\boldsymbol{r}_{3}\right)=\rho_{0}\exp\left\{ -\frac{1}{b^{2}}\left[\left(\sum_{i=1}^{3}\boldsymbol{r}_{i}^{2}\right)+\left(\sum_{i=1}^{3}\boldsymbol{r}_{i}\right)^{2}\right]\right\} ,\label{eq:gauss-gs-density}
	\end{equation}
	with normalization \(4\pi\int_{0}^{\infty}\rho\left(r\right)r^{2}dr=4\) and Gaussian parameter \(b=1.41\) fm~\cite{DEJAGER1974479}. By straightforward application of Eqs. \eqref{eq:U_rho}, \eqref{eq:gs-3r}, and Eq. \eqref{eq:gauss-gs-density}, the potential can be written as
	\begin{equation}
		U\left(r\right) = -\frac{32}{\sqrt{3\pi}}\frac{a_{0}}{b^{3}}\exp\left(-\frac{4r^{2}}{3b^{2}}\right).\label{eq:optical_pot_final}
	\end{equation}

	\subsection{Choice of N-core interaction}
	Two types of N$\alpha$ interactions were tried. Both fit N$\alpha$ scattering phase shifts~\cite{RevModPhys.57.923} satisfactorily.
	The first, SBB $V_{n\alpha}$ potential~\cite{Sack-PhysRev.93.321} has a Gaussian form given by
	\begin{equation}
		V_{n\alpha}\left(r\right)=\sum_{l=0,1,2}v^{\left(l\right)}\exp\left[-\left(r/b\right)^{2}\right]
		+ v^{\left(ls\right)}\exp\left[-\left(r/b\right)^{2}\right],
		\label{eq:sbb}
	\end{equation}
	having $s$-, $p$-, $d$- and $l$s components of different strengths but of a common range $b=2.35$ fm.
	The strength parameters of the original SBB potential in MeV for different components are $v^{\left(s\right)}=50.0,v^{\left(p\right)}=47.32,v^{\left(d\right)}=-23.0$
	and $v^{\left(ls\right)}=-11.71$ respectively. 
	The repulsive s-component in the core$(\alpha)$-N potential has been introduced to simulate
	Pauli principle between valence neutrons and nucleons forming the core nuclei.
	
	The second choice $V_{n\alpha}$(WS) employs a Woods-Saxon form for
	the corresponding potential components as used in the coordinate space
	Faddeev calculation~\cite{BANG1983126,Zhukov93}
	for $^{6}\textrm{Li}$, and from Refs.~\cite{Thompson-prc-2000,casal2013},
	with central $V_{c}\left(r\right)$ and spin-orbit $V_{so}\left(r\right)$
	terms which are expressed as Woods-Saxon potentials,
	\begin{equation}
		V_{c}\left(r\right)=\frac{v_{c}^{\left(l\right)}}{1+\exp\left(\frac{r-r_{c}^{\left(l\right)}}{a_{c}^{\left(l\right)}}\right)}, \label{eq:ws-vc}
	\end{equation}
	
	\begin{equation}
		V_{so}\left(r\right)=\frac{v_{so}}{ra_{so}}\frac{\exp\left(\frac{r-r_{so}}{a_{so}}\right)}{\left[1+\exp\left(\frac{r-r_{so}}{a_{so}}\right)\right]^{2}}. \label{eq:ws-vso}
	\end{equation}
	Again, the parameters for the central part are $l$-dependent, $v_{c}^{\left(s\right)}=48.0,v_{c}^{\left(p\right)}=-43.0,v_{c}^{\left(d\right)}=-21.5$
	MeV with same $r_{c}^{\left(s, p, d\right)}=2.0$ fm and $a_{c}^{\left(s, p, d\right)}= 0.7 $ fm. 
	The spin-orbit potential parameters are given by: $v_{so}=-40.0$ MeV $\textrm{fm}^{2}$, $r_{so}=1.5$ fm and $a_{so}=0.35$ fm.
	
	%%%%%%%%%%%%%%%	
	\section{Numerical results} 
	\label{result}		
    The coupled equations, Eqs.~\eqref{eq:Faddeev_coupled} were solved using the FaCE computational toolkit~\cite{face}, with the two-body interactions described in Sec.~\ref{sec:Two-body-potentials}. 	
	In the HH method, the final results are dependent on the maximum hypermomentum value, $K_{max}$, due to the truncation of the total three-body wave function expansion in hypermomentum components. Therefore, an initial investigation into the convergence of results was performed as a function of $K_{max}$ and the maximum number of hyperradial excitations, $i_{max}$. For the $\phi\textrm{N-}\alpha$ state, convergence was achieved satisfactorily with $K_{max} = 70$ and $i_{max} = 25$.
	
	The HAL QCD $\phi$N two-body potential in the $^{4}S_{3/2}$ channel (Eq.~\eqref{eq:phiN-pot} with parameters provided in Table~\ref{tab:Fit_para}) was found not to form a bound state. In contrast, the $\phi$N interaction in the $^{2}S_{1/2}$ channel results in a strongly bound state, with central binding energies of $28.93$ MeV and $16.64$ MeV at lattice Euclidean times $t/a = 12$~\cite{CHIZZALI2024138358} and $14$~\cite{yan2022prd}, respectively.

\begin{figure*}[hbt!]
		\centering
		\includegraphics[scale=1.0]{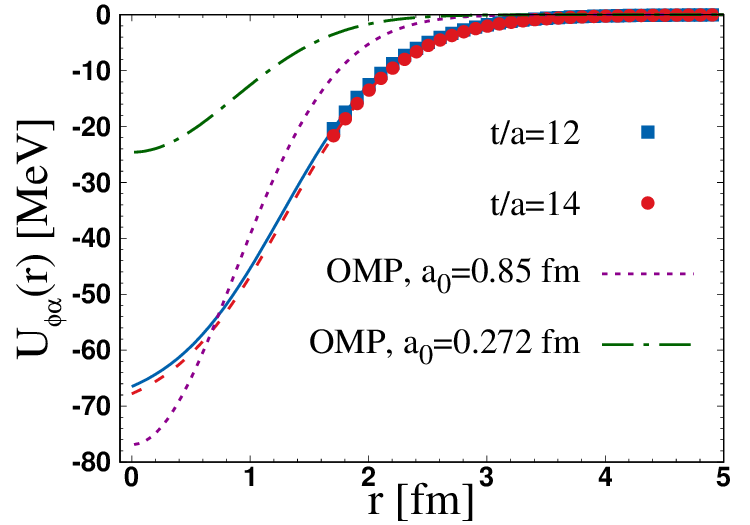}
		\caption{
			The  spin-averaged $ \phi \alpha $ potential in Eq.~\eqref{eq:ws-fit} as a function of separation $ r $ for  lattice Euclidean time $t/a= 12 $ (blue solid line), and $ 14 $ (red dashed lines) by the parameters given in Table~\ref{tab:phi-alpha-para}.  
			The $\phi \alpha$ potential, obtained from the OMP approximation as given by Eq.~\eqref{eq:optical_pot_final}, is based on $a_{0} = 0.85$ fm (dotted purple lines) extracted by the ALICE collaboration in~\cite{PhysRevLett.127.172301} and on $a_{0} = 0.272$ fm (dash-dotted green lines) obtained from the theoretical coupled-channel approach within the chiral Lagrangian method \cite{FeijooPRD2025}.		
		}
		\label{fig:alphaphi_pot}
	\end{figure*}

	The central binding energy for the $\phi\alpha$ system, \( B_{\phi\alpha} \), and the corresponding parameters are given in Table~\ref{tab:phi-alpha-para}. The $\phi\alpha$ potentials obtained via SFP (Eq.~\eqref{eq:ws-fit}) and OMP (Eq.~\eqref{eq:optical_pot_final}) methods are depicted in Fig.~\ref{fig:alphaphi_pot}. Calculations are performed with the experimental masses, \( m_{\alpha} = 3727.38 \) MeV/\( c^{2} \) and \( m_{\phi} = 1019.5 \) MeV/\( c^{2} \). The central binding energy of the $\phi\alpha$ system from the OMP approximation is calculated using the corresponding scattering lengths, \( a_{0} = 0.850 \) fm for purely elastic~\cite{PhysRevLett.127.172301} and \( a_{0} = 0.272 \) fm for a coupled-channel~\cite{FeijooPRD2025} $\phi p$ interactions; for the latter, no bound state is found.
	
   Moreover, from Fig.~1 in Ref.~\cite{FeijooPRD2025}, it is seen that the $\phi p$ CF for the pure theoretical model is somewhat larger than that of the bootstrap model (which aligns most closely with experimental measurements). Consequently, it may be concluded that the $\phi p$ interaction in the bootstrap model coming from the resampling of the data, with \( a_{0} = -0.034 \pm 0.035  \) fm, is less attractive than the interaction in the pure theoretical model with \( a_{0} = 0.850 \) fm. Therefore, it can be concluded that the $\phi p$ interaction in the bootstrap model does not lead to a bound state in the $\phi \alpha$ system.
	
		%%%%%%%%%%%%%%%%
	\begin{table}[hbt!]
		\caption{
			The fit parameters for $U_{\phi\alpha}$ in Eq.~\eqref{eq:ws-fit} are presented in conjunction with their corresponding central binding energy, $B_{\phi\alpha}$. The fitting range is selected for $r \gtrsim 1.70$ fm~\cite{filikhin2024folding}. In Ref.~\cite{filikhin2024phihe}, a $B_{\phi\alpha}$ value of 4.78 MeV is reported, with the $\phi \alpha$ potential is constructed based solely on the $\phi\alpha$ interaction in the $^{4}S_{3/2}$ channel, and the rms radius set at 1.70 fm.
			$B_{\phi\alpha}$ from the OMP approximation is obtained employing $a_{0}$ for pure elastic scattering \cite{PhysRevLett.127.172301} and the pure theoretical coupled-channels $\phi p$ interaction \cite{FeijooPRD2025}; for the latter, no bound (NB) state is found.
			\label{tab:phi-alpha-para}}	
		\begin{tabular}{ccccc}
			\hline
			\hline 
			t/a & $ U_{0} $ (MeV)& $ R_{\phi\alpha} $(fm)& $ t_{\phi\alpha} $ (fm)& $B_{\phi\alpha}$(MeV) \\
			\hline
			$ 12 $& $ 71.1 $ & $ 1.270 $ & $ 0.476 $ &  $ 10.29 $ \\	
			$ 14 $& $ 72.5 $ & $ 1.290 $ & $ 0.485 $ &  $ 11.38 $ \\
			\hline
			\multirow{3}{*}{OMP}&& $ a_{0} $(fm) &                  &$B_{\phi\alpha}$(MeV)\\ \cline{3-5}  
			&& $ 0.850 $~\cite{PhysRevLett.127.172301} &&  $ 5.30 $ \\
			&& $ 0.272 $~\cite{FeijooPRD2025}          &&  NB \\
			\hline 	
		\end{tabular}
	\end{table}
	%%%%%%%%
	%-7.1114138718299643e+01     4.7596960982880659e-01     1.2692502865129618e+00   -10.289062500000000
	%-7.2541279172057457e+01     4.8524394544015709e-01     1.2886502875352259e+00   -11.378906250000000
	%%%%%%%%%%%%%%%%%%%%%%%%
	%%%%%%%%%%%%%%	
	%%%%%%%%%%%

	The ground state central binding energies and nuclear matter radii of $\phi\textrm{N-}\alpha$ mesic nuclei within the $^{4}S_{3/2}$ and $^{2}S_{1/2}$ spin channels are provided in Table~\ref{tab:pot3}.	
	The central binding energies of the $ \phi\textrm{N-}\alpha $ bound states in the spin $ 3/2\left(1/2\right) $ channel are found to be about $ 10 \left(25\right) $ and $ 11 \left(13\right) $ MeV at Euclidean times $ t/a=12 $ and $14$, respectively.
	A significant discrepancy has been observed between the binding energies of the $ \phi\textrm{N-}\alpha \left(^{2}S_{1/2}\right)$ state at Euclidean times $ t/a=12 $ ($25$ MeV) and $ t/a=14 $ ($13$ MeV). This difference is directly associated with the binding energies of the two-body $\phi$N system at $ t/a=12 $ and $14$, which are approximately 29 MeV and 17 MeV, respectively, as noted in the preceding paragraph.

	Also, the results with using the obtained  $\phi\alpha$ potentials through the OMP (Eq.~\eqref{eq:optical_pot_final}) methods based on the pure elastic single~\cite{PhysRevLett.127.172301} ($ a_{0} = 0.850 $ fm) and the pure theoretical coupled-channel~\cite{FeijooPRD2025} ($ a_{0} = 0.272$ fm) $\phi$N interactions, are provided in Table~\ref{tab:pot3}.
	 For the latter case, no bound state is found, even though the spin-averaged $\phi$N potential (Eq.~\eqref{eq:spin-ave}) is used, which is found to be more attractive than the value calculated in Ref.~\cite{FeijooPRD2025}. 

	Furthermore, in order to compute the root-mean-square (rms) matter radius of the $ \phi\textrm{N-}\alpha $ state, the rms matter radius of the $\alpha$ particle, taken as 1.7 fm~\cite{PhysRevC.109.L012201}, along with the strong interaction radius of the neutron ($0.80$ fm) and the $\phi$-meson ($0.46$ fm)~\cite{PhysRevC.108.034614,POVH1990653}, has been utilized in the calculation. The resulting nuclear matter radii of the $ \phi\textrm{N-}\alpha $ bound states in the spin $ 3/2 \left(1/2\right) $ channels are estimated to be approximately 4.5 (1.7) fm  and 4.6 (1.8) fm at Euclidean times $ t/a=12 $ and $ t/a=14 $, respectively.		
	%
	%%%%%%%%%%%%%%%%%%%%%%%%
	\begin{table}
		\caption{Three-body ground state central binding energies $\left(B_{3}\right)$ and the nuclear matter radii $ \left(r_{\textrm{mat}}\right) $ of the $ \phi\textrm{N-}\alpha $ system for $ \phi $N potentials in $ ^{4}S_{3/2} $ and $^{2}S_{1/2}$ channels,  and two types of N$\alpha$ interactions, i.e., the SBB potential that is given by Eq.~\eqref{eq:sbb} and WS potential which is set by Eqs.~\eqref{eq:ws-vc},~\eqref{eq:ws-vso}.
			It is worth recalling that the OMP shows that calculations using $\phi\alpha$ potentials based on the elastic single~\cite{PhysRevLett.127.172301} ($ a_{0} = 0.850 $ fm) and the pure theoretical coupled-channels~\cite{FeijooPRD2025} ($ a_{0} = 0.272$ fm) $\phi$N interactions.
			%%%%			
			 \label{tab:pot3}  
		}
		\centering
		%	\begin{ruledtabular}
			\begin{tabular}{ccccccc}
				\hline \hline
				$\textrm{N}\alpha:$	&	 &  \multicolumn{2}{c}{$ \textrm{SBB} $}    && \multicolumn{2}{c}{$\textrm{WS}$}    \\ \cline{3-4} \cline{6-7} 
				$t/a$&$ \phi $N spin &   $B_3$ (MeV) & $r_{mat}$ (fm) && $B_3$ (MeV) & $r_{mat}$ (fm) \\
				\hline  
				\multirow{2}{*}{$12$} &$ ^{4}S_{3/2} $ & $ 9.68 $  & $ 4.55 $ && $ 8.70 $ & $ 3.45 $  \\
				&$ ^{2}S_{1/2} $ & $ 25.21 $ & $ 1.71 $ && $ 25.10$ & $ 1.70 $  \\
				\hline
				\multirow{2}{*}{$14$} &$ ^{4}S_{3/2} $ & $ 10.80 $ & $ 4.59 $ && $  9.84 $ & $ 3.45 $   \\
				&$ ^{2}S_{1/2} $ & $ 13.36 $ & $ 1.79 $ && $ 13.32 $ & $ 1.78 $   \\
				\hline
				\multirow{3}{*}{OMP}  & $ a_{0} $(fm)  &         &          &&   &  \\ \cline{2-2}  
			& $ 0.850 $~\cite{PhysRevLett.127.172301}  &  $3.28$ & $3.40$   && $ 3.27 $ & $ 3.40 $\\
            & $ 0.272 $~\cite{FeijooPRD2025}           &   NB     &  -       && NB & -\\
            \hline \hline
			\end{tabular}
			%\end{ruledtabular}
			
		\end{table}
		%%%%%%%%%%%%%%%%%%%%%%%%
		%& $ 9.679675 $  & $ 4.549 $ && $ 8.704484 $ & $ 3.448 $\\& $ 25.212638 $ & $ 1.707 $ && $ 25.095992 $ & $ 1.702 $  \\			
		%& $ 10.792995 $ & $ 4.590 $ && $  9.839589 $ & $ 3.455 $   \\	&$ ^{2}S_{1/2} $ & $ 13.361447 $ & $ 1.786 $ && $ 13.319550 $ & $ 1.781 $		

		\section{Summary and conclusions\label{sec:Summary-and-conclusions}}
		The modern $ \phi $N interactions that is obtained from the analysis of the pure elastic scattering and the coupled-channels in the $\phi p$ correlation functions, has been investigated through the analysis of bound states in the $\phi \textrm{N-} \alpha$ system. In this approach, the $\alpha$ cluster is utilized to attract the $\phi$N pair without altering its spin configuration. 
		The $ \phi\alpha $ potentials are constructed through two methods, SFP method for given spin-averaged $ \phi $N potentials in coordinate space and the OMP approximation within the multiple-scattering framework for given scattering length of the $ \phi $N interaction. The N$\alpha$ interaction potential is modeled using two commonly adopted forms from the literature: one with a Gaussian profile (SBB) and the other with a Woods-Saxon profile, both including central and spin-orbit components.  
		 Subsequently, the coupled Faddeev equations in coordinate space are solved within the HH expansion method, using the two-body potentials for each particle (cluster) pair.

		The numerical results showed that using only the single channel $ \phi $N interactions, $ \phi\textrm{N-}\alpha $ system could be bound with a binding energy within $\left[ 3.3-25.2\right]$ MeV.  As well as,  the corresponding nuclear matter radii were estimated ranging from 1.70 to 4.59 fm.
		But, in the case of the coupled-channel $\phi p$ interaction, which is the most consistent with experimental measurements~\cite{FeijooPRD2025}, 	
		 no bound state is found, even though the spin-averaged $\phi$N potential is used, which is found to be more attractive than the coupled-channel counterpart. 
		 It is noteworthy that, when the $\phi p$ interaction in the bootstrap model, with \( a_{0} = -0.034 \pm 0.035 \) fm, which arises from data resampling, is used, a bound state is very far from being obtained.
		In conclusion, the present results show that,  it is essential to consider the contributions from the dynamics of the vector baryon coupled channels $\phi p$ interaction. This effect could play a decisive role in the existence of meson nuclei.

		Lattice calculations of the $\phi$N interaction in the $^{4}S_{3/2}$ channel are used in~\cite{CHIZZALI2024138358} to constrain the spin-$1/2$ ($^{2}S_{1/2}$) channel by fitting the experimental $\phi$N correlation function measured by the ALICE Collaboration~\cite{PhysRevLett.127.172301}. For a meaningful comparison between theory and experiment, it is necessary to extract complex-valued scattering parameters in both the $^{4}S_{3/2}$ and $^{2}S_{1/2}$ channels through a coupled-channels analysis~\cite{sasaki2020,etminan2024prd} of the data from physical-point simulations~\cite{ScalePhysRevD.110.094502}. It is hoped that the presented numerical results will aid in the planning of future experiments.

		%\section*{Acknowledgement}				
		%%%%%%%%%%%				
		%\section*{References}
		\bibliography{Refs.bib}
	\end{document}